\documentclass[aps,twocolumn,prd,epsf]{revtex4}
\usepackage{amsfonts}
\usepackage{amssymb}
\usepackage{graphicx}

\newcommand{\be}{\begin{equation}}
\newcommand{\ee}{\end{equation}}
\newcommand{\bea}{\begin{eqnarray}}
\newcommand{\eea}{\end{eqnarray}}

\begin{document}

\title{Warped product spaces and geodesic motion in the neighbourhood of
hypersurfaces}
\author{F. Dahia$^{a}$, C. Romero$^{b}$, L. F. P. da Silva,$^{b}$ and R.
Tavakol$^{c}$}

\begin{abstract}
We study the classical geodesic motions of nonzero rest mass test particles
and photons in five-dimensional warped product spaces. We show that it is
possible to obtain a general picture of these motions, using the natural
decoupling that occurs in such spaces between the motions in the fifth
dimension and the motion in the hypersurfaces. This splitting allows the use
of phase space analysis in order to investigate the possible confinement of
particles and photons to hypersurfaces in five-dimensional warped product
spaces. Using such analysis, we find a novel form of quasi-confinement which
is oscillatory and neutrally stable. We also find that this class of warped
product spaces \emph{locally} satisfy the $Z_{2}$ symmetry by default. The
importance of such a confinement is that it is purely due to the classical
gravitational effects, without requiring the presence of brane--type
confinement mechanisms. \newline

%
%
%
%
\newline
\end{abstract}


\address{$^{a}$Departamento de F\'{\i}sica, Universidade Federal \\
de Campina Grande, 58109-970\\
Campina Grande, Pb, Brazil\\
$^{b}$Departamento de F\'{\i}sica, Universidade Federal da Para\'{\i}ba, \\
Caixa Postal 5008, 58059-979 Jo\~{a}o Pessoa, Pb, Brazil\\
$^{c}$School of Mathematical Sciences , Queen Mary, University of London,\\
London E1 4NS United Kingdom }


\maketitle

\section{Introduction}

Motivated by string inspired theories \cite{Arkani,Randall}, a great deal of
effort has, over the recent years, gone into the study of higher dimensional
general relativity (GR) theories. Examples of such theories include
noncompactified $(n+1)$-dimensional GR \cite{ndGR} (where the observable
universe is assumed to lie on an embedded $3+1$ (4D) hypersurface),
Kaluza--Klein theory \cite{KK} (where the extra dimensions are compactified)
and the braneworld models inspired by String/M-theory \cite%
{Randall,SMS,Maartensrev}.

A central issue in such theories has been how to explain the fact that the
observable universe is (very nearly) confined to a particular hypersurface.
In most braneworld models, stringy effects are invoked to argue that in low
energy regimes particles are restricted to a special $3+1$ \textit{brane}
hypersurface, which is embedded in a higher (usually five)-dimensional
\textit{bulk}, while the gravitational field is free to propagate in the
bulk \cite{Arkani}.

In the case of other higher-dimensional GR theories, however, such
confinement needs to be explained. This has provided strong motivation for
the consideration of the so-called warped product spaces \cite{Frolov} and
the investigation of their geometrical properties. We recall that the
geometrical considerations of warped product spacetimes predate their
applications in these higher-dimensional contexts and were first considered
by Bishop and O\'{ }Neill \cite{Bishop} in the context of $3+1$ general
relativity.

Geodesic motion in connection with higher-dimensional theories has been
considered by a number of authors in recent years \cite{Wesson1,Seahra,Muck}%
. Our interest in studying such motions specifically in the context of
warped product spaces arises from the realization of the fact that there is
a strong connection between the 5D geodesics and the analytical form of the
warping function. Indeed, as we shall see in this paper, it is possible to
make a general qualitative analysis of the behaviour of massive particles
and photons in the fifth dimension from the knowledge of the warping
function. This can in turn allow questions concerning the possibility of
hypersurface confinement as well as the stability of the confined motions
with respect to small perturbations to be treated with great generality.

An assumption usually made in the braneworld scenarios is that the branes
possess an infinitesimal thickness. In this sense one may talk about \textit{%
$\delta $-confinement} of massive particles and light rays. There are also
smooth versions of this scenario which consider the possibility that branes
may have a finite thickness \cite{Csaki,Gremm,Wolfe}. In that case one may
talk about \textit{quasi-confinement}.

Here we shall demonstrate a novel example of the latter in the context of $%
4+1$-dimensional warped product spaces which we shall refer to as \textit{%
oscillatory confinement}, whereby the particles in the four-dimensional
hypersurface can oscillate about the hypersurface while remaining close to
it. As we shall show below such behaviour can indeed occur for large classes
of bulks which possess warped product geometries. Our analysis is fairly
general and the results may be applied to different examples of warped
product spacetimes considered in the literature.

The paper is organized as follows. In Section 2 we write the geodesic
equations for warped product spaces and consider the parts due to 4D and the
fifth dimension separately. We then show that the equation that describes
the motion in 5D decouples from the rest. We proceed in Section 3 to rewrite
the geodesic equation in the fifth dimension as an autonomous planar
dynamical system. We then employ phase plane analysis to study the motion of
particles with nonzero rest mass (timelike geodesics) and photons (null
geodesics) respectively. Such qualitative analysis allows very general
conclusions to be drawn about the possible existence of confined motions and
their stability in the neighbourhood of hypersurfaces. We conclude with a
discussion of further applications in Section 4.

Throughout Latin indices take values in the range (0,1,...,4) while Greek
indices run over (0,1,2,3).

\section{Warped product spaces and five-dimensional geodesics}

Warped geometries are at the core of the Randall-Sundrum models. Technically
we can define a warped product space in the following way. Let $(M^{m},h)$
and $(M^{n},k)$ be two\ Riemannian (or pseudo-Riemannian) manifolds of
dimension $m$ and $n$, with metrics $h$ and $k,$ respectively. Suppose we
are given a smooth function $f:M^{n}\rightarrow
\mathbb{R}
$ (which is referred to as the \textit{warping function}). Then we can
construct a new Riemannian (pseudo-Riemannian) manifold by setting $%
M=M^{n}\times M^{m}$ and defining a metric $g=e^{2f}h\otimes k$. In this
paper we shall take $M=M^{4}\times \mathbb{R}$, where $M^{4}$ is a
four-dimensional Lorentzian manifold with signature $(+---)$ (referred to as
the (3+1)-dimensional \textit{spacetime}) . In local coordinates $\{y^{a}\}$
the line element corresponding to this metric will be denoted by
\[
dS^{2}=g_{ab}dy^{a}dy^{b}.
\]

Let us now consider the equations of geodesics in the five-dimensional space
\begin{equation}
\frac{d^{2}y^{a}}{d\lambda ^{2}}+^{(5)}\Gamma _{bc}^{a}\frac{dy^{b}}{%
d\lambda }\frac{dy^{c}}{d\lambda }=0,  \label{geodesics5D}
\end{equation}%
where $\lambda $ is an affine parameter and $^{(5)}\Gamma _{bc}^{a}$ are the
5D Christoffel symbols of the second kind defined by $^{(5)}\Gamma _{bc}^{a}=%
\frac{1}{2}g^{ad}\left( g_{db,c}+g_{dc,b}-g_{bc,d}\right) $. Denoting the
fifth coordinate $y^{4}$ by $l$ and the remaining coordinates $y^{\mu }$\
(the "spacetime" \textit{\ }coordinates) by $x^{\mu }$, i.e. $y^{a}=(x^{\mu
},l)$, it is not difficult to show that the "4D part" of the geodesic
equations (\ref{geodesics5D}) can be rewritten in the form
\begin{equation}
\frac{d^{2}x^{\mu }}{d\lambda ^{2}}+^{(4)}\Gamma _{\alpha \beta }^{\mu }%
\frac{dx^{\alpha }}{d\lambda }\frac{dx^{\beta }}{d\lambda }=\phi ^{\mu },
\label{4Dpart}
\end{equation}%
where
\begin{eqnarray}
\phi ^{\mu } &=&-^{(5)}\Gamma _{44}^{\mu }\left( \frac{dl}{d\lambda }\right)
^{2}-2^{(5)}\Gamma _{\alpha 4}^{\mu }\frac{dx^{\alpha }}{d\lambda }\frac{dl}{%
d\lambda }  \nonumber \\
&-&\frac{1}{2}g^{\mu 4}\left( g_{4\alpha ,\beta }+g_{4\beta ,\alpha
}-g_{\alpha \beta ,4}\right) \frac{dx^{\alpha }}{d\lambda }\frac{dx^{\beta }%
}{d\lambda },
\end{eqnarray}%
and $^{(4)}\Gamma _{\alpha \beta }^{\mu }=\frac{1}{2}g^{\mu \nu }\left(
g_{\nu \alpha ,\beta }+g_{\nu \beta ,\alpha }-g_{\alpha \beta ,\nu }\right) $%
. We shall assume that the five-dimensional manifold $M$ can be foliated by
a family of hypersurfaces $\Sigma $ defined by the equation $l=$ constant.
Then the geometry of each hypersurface, say \ $l=l_{0},$ will be determined
by the induced metric
\[
ds^{2}=g_{\alpha \beta }(x,l_{0})dx^{\alpha }dx^{\beta }.
\]%
Therefore the quantities $^{(4)}\Gamma _{\alpha \beta }^{\mu }$ which appear
on the left-hand side of Eq. (\ref{4Dpart}) may be identified with the
Christoffel symbols associated with the induced metric in the leaves of the
foliation defined above.

We shall consider the class of warped geometries given by the following line
element
\begin{equation}
dS^{2}=e^{2f}h_{\alpha \beta }dx^{\alpha }dx^{\beta }-dl^{2},  \label{warped}
\end{equation}%
where $f=f(l)$ and $h_{\alpha \beta }=$ $h_{\alpha \beta }(x)$. For this
metric it is easy to see \footnote{%
In the above calculation we have used the fact that the matrix $h_{\alpha
\beta }$ has an inverse $h^{\alpha \beta }$, that is, $\ h^{\mu \beta
}h_{\beta \nu }=\delta _{\upsilon }^{\mu }$. This may be easily seen since,
by definition, $\det h=-\det g\neq 0$.} that $^{(5)}\Gamma _{44}^{\mu }=0$
and $^{(5)}\Gamma _{4\nu }^{\mu }=\frac{1}{2}g^{\mu \beta }g_{\beta \nu
,4}=f^{\prime }\delta _{\nu }^{\mu }$, where a prime denotes a derivative
with respect to $l$. Thus in the case of the warped product space (\ref%
{warped}) the right-hand side of Eq. (\ref{4Dpart}) reduces to $\phi ^{\mu
}=-2f^{\prime }\frac{dx^{\mu }}{d\lambda }\frac{dl}{d\lambda }$ and the 4D
part of the geodesic equations becomes
\begin{equation}
\frac{d^{2}x^{\mu }}{d\lambda ^{2}}+^{(4)}\Gamma _{\alpha \beta }^{\mu }%
\frac{dx^{\alpha }}{d\lambda }\frac{dx^{\beta }}{d\lambda }=-2f^{\prime }%
\frac{dx^{\mu }}{d\lambda }\frac{dl}{d\lambda }.  \label{4Dwarped}
\end{equation}%
Likewise the geodesic equation for the fifth coordinate $l$ in the warped
product space becomes

\begin{equation}
\frac{d^{2}l}{d\lambda ^{2}}+f^{\prime }e^{2f}h_{\alpha \beta }\frac{%
dx^{\alpha }}{d\lambda }\frac{dx^{\beta }}{d\lambda }=0.  \label{lwarped}
\end{equation}%
By restricting ourselves to 5D timelike geodesics $\left( g_{ab}\frac{dy^{a}%
}{d\lambda }\frac{dy^{b}}{d\lambda }=1\right) $ we can readily decouple the
above equation from the 4D spacetime coordinates to obtain
\begin{equation}
\frac{d^{2}l}{d\lambda ^{2}}+f^{\prime }\left( 1+\left( \frac{dl}{d\lambda }%
\right) ^{2}\right) =0.  \label{5Dmotion}
\end{equation}%
Similarly, to study the motion of photons in 5D, we must consider the null
geodesics $\left( g_{ab}\frac{dy^{a}}{d\lambda }\frac{dy^{b}}{d\lambda }%
=0\right) $, in which case Eq. (\ref{lwarped}) becomes
\begin{equation}
\frac{d^{2}l}{d\lambda ^{2}}+f^{\prime }\left( \frac{dl}{d\lambda }\right)
^{2}=0.  \label{5Dphoton}
\end{equation}

\bigskip Equations (\ref{5Dmotion}) and (\ref{5Dphoton}) are ordinary
differential equations of second-order which, in principle, can be solved if
the function $f^{\prime }=f^{\prime }(l)$ is given. Conversely, if $%
l=l(\lambda )$ is known then $f=f(l)$ can be determined (up to a constant of
integration) provided that we can write $\lambda =\lambda (l)$.

\section{Qualitative analysis of motion in the fifth dimension}

We recall that the hypersurface sets the boundary conditions for the warping
function. If the warping function $f=f(l)$ is given a priori, then we can
obtain a great deal of information about the motion in the fifth dimension
by performing a qualitative analysis of Eqs. (\ref{5Dmotion})\ and (\ref%
{5Dphoton}), without the need to solve these equations analytically. To do
this we define the variable $q=\frac{dl}{d\lambda }$ and consider instead of
(\ref{5Dmotion}) or (\ref{5Dphoton})\ the autonomous dynamical system \cite%
{Andronov}
\begin{eqnarray}  \label{dynamical}
\frac{dl}{d\lambda}&=&q  \nonumber \\
\frac{dq}{d\lambda}&=&P(q,l)
\end{eqnarray}
with $P(q,l)=-f^{\prime }(\epsilon +q^{2})$, where$\ \epsilon =1$ in the
case of (\ref{5Dmotion}) (corresponding to the motion of particles with
nonzero rest mass) and $\epsilon =0$ in the case of (\ref{5Dphoton})
(corresponding to the motion of photons). In the investigation of dynamical
systems a crucial role is played by their \textit{equilibrium points}, which
in the case of system (\ref{dynamical}) are given by $\frac{dl}{d\lambda }=0=%
\frac{dq}{d\lambda }$. The knowledge of these points together with their
stability properties allows a great deal of information to be gained
regarding the types of behaviour allowed by the system. Thus if for a
particular geodesic we know $t$ as a function of the parameter $\lambda $,
i.e. $t=\ t(\lambda )$, then since $\frac{dt}{d\lambda }\neq 0$, knowing the
behaviour of $l=l(\lambda )$ from qualitative analysis will enable us to
deduce the time evolution of the motion of a particle (or light ray) in the
fifth dimension.

\subsection{The 5D motion of particles with nonzero rest mass near the
hypersurface}

We begin by considering the case of nonzero rest mass particles, whose
motion in the fifth dimension is governed by the dynamical system
\begin{eqnarray}  \label{dynamical1}
\frac{dl}{d\lambda }& = &q  \nonumber \\
\frac{dq}{d\lambda }& = & -f^{\prime }(1+q^{2})
\end{eqnarray}
The equilibrium points of (\ref{dynamical1}) are given by $q=0$ and the
zeros of the function $f^{\prime }(l)$ (if they exist) which we generically
denote by $l_{0}$. These solutions, pictured as fixed points in the phase
plane, correspond to curves which lie entirely in a hypersurface $\Sigma $
of our foliation (since they have $l=$ constant). It turns out that these
curves are timelike geodesics with respect to the hypersurface induced
geometry. Indeed, the existence of equilibrium points of the dynamical
system (\ref{dynamical}) has the important consequence that the geodesics of
the bulk and that of the hypersurface ($l_{0}=$ constant) coincide. This can
be seen directly from the equations (\ref{4Dpart}) and (\ref{lwarped}), and
is perfectly consistent with a well-known theorem of differential geometry
according to which the geodesics of a Riemannian space and that of a
particular hypersurface coincide (i.e. confinement occurs) if and only if
the extrinsic curvature of that hypersurface vanishes (see e.g. \cite%
{Eisenhart} where such hypersurfaces are referred to as \textit{totally
geodesic}). Note that for the metric (\ref{warped}) the extrinsic curvature
of the hypersurfaces $\Sigma $ is given by $\Omega _{\alpha \beta
}=-f^{\prime }e^{2f}h_{\alpha \beta }(x)$, which is clearly zero at the
equilibrium points, where $f^{\prime }=0$. Thus in the absence of
equilibrium points (i.e. $f^{\prime }\neq 0$) the extrinsic curvature does
not vanish and according to the above theorem the bulk and hypersurface
geodesics cannot be identical.

To obtain information about the possible modes of behaviour of particles and
light rays in such hypersurfaces, it is important to study the nature and
stability of the corresponding equilibrium points. This can be done by
linearising equations (\ref{dynamical1}) and studying the eigenvalues of the
corresponding Jacobian matrix about the equilibrium points. Assuming that
the function $f^{\prime }(l)$ vanishes, at least at one point $l_{0}$, it
can readily be shown that the corresponding eigenvalues are determined by
the sign of the second derivative $f^{\prime \prime }(l_{0})$, at the
equilibrium point, and the following possibilities can arise for the
equilibrium points of the dynamical system (\ref{dynamical1}): \newline

\begin{figure}[tbp]
\caption{Depicted in the top panel is the center equilibrium point E in the
case where $f^{\prime \prime }(l_{0})>0$. In this case the particles
oscillate about the hypersurface $l=l_{0}$. This provides a
quasi-confinement mechanism whereby massive particles enter and leave the
hypersurface $\Sigma $ indefinitely (lower panel).}
\label{Fig1}\includegraphics[height=3.0in,width=3.0in]{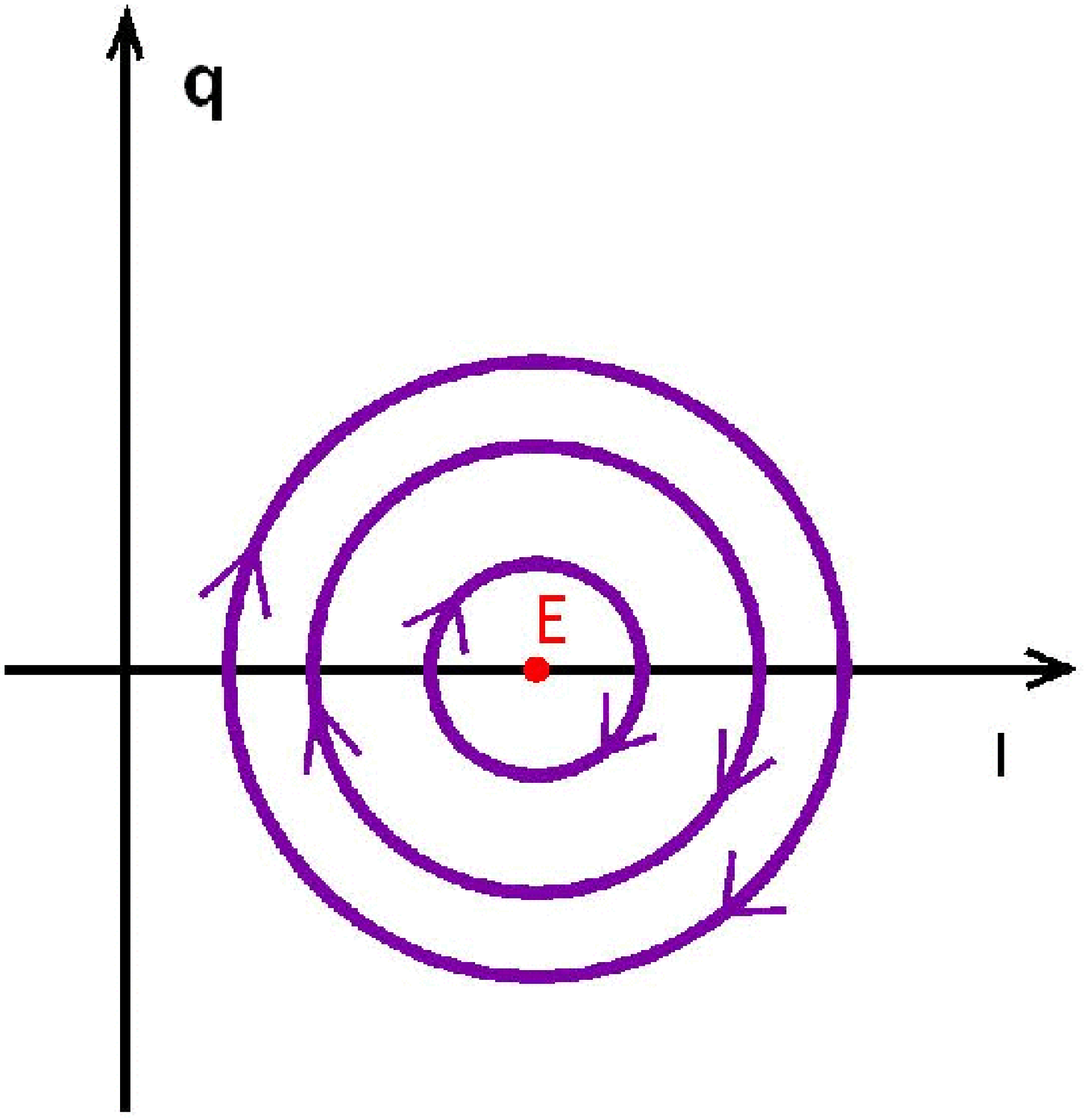} %
\includegraphics[height=3.0in,width=3.0in]{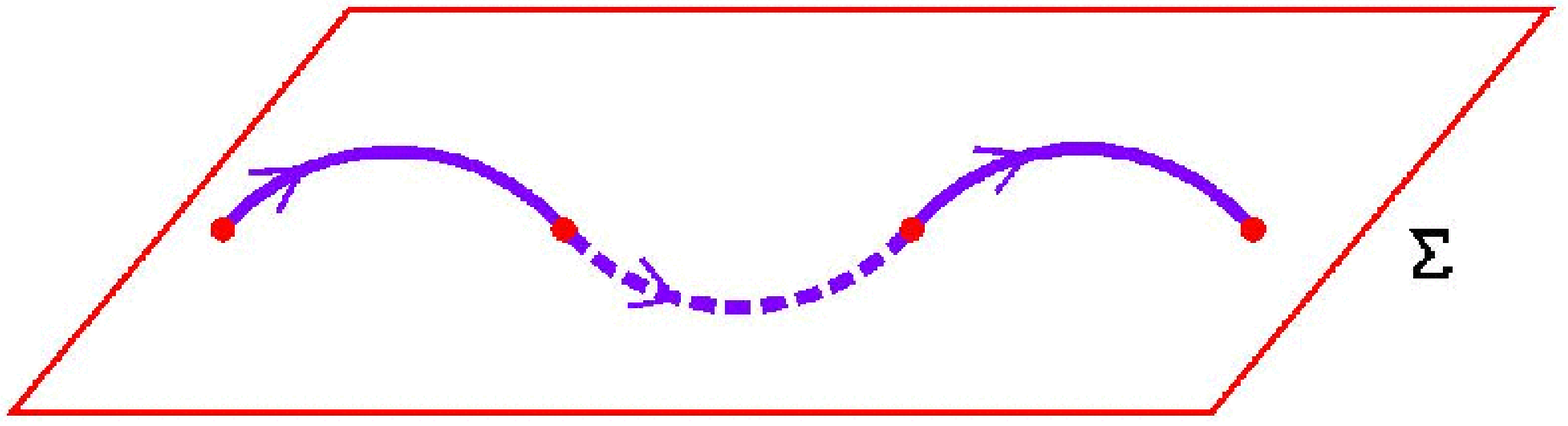}
\end{figure}

Case I. If $f^{\prime \prime }(l_{0})>0$, then the equilibrium point $%
(q=0,l=l_{0})$ is a \textit{center}. This represents a case in which the
solutions near the equilibrium point have the topology of a circle. In that
case the phase portrait consists of closed curves describing the motions of
particles oscillating about the hypersurface $\Sigma$ ($l=l_{0}$)
indefinitely (see the top panel in Fig.~\ref{Fig1}). The amplitudes of the
oscillations will depend on initial conditions. We note that the existence
of such cyclic motions is independent of the ordinary 4D spacetime
dimensions, and, except for the conditions $f^{\prime }(l_{0})=0$ and\ $%
f^{\prime \prime }(l_{0})>0$, the warping function $f(l)$ remains completely
arbitrary. The presence of such center equilibrium points with the
consequent quasi-confinement of particles can be viewed as providing
examples of \textit{almost totally geodesic hypersurfaces } (see the lower
panel in Fig.~\ref{Fig1}).
\newline

Case II. If $f^{\prime \prime }(l_{0})<0$, then the point $(q=0,l=l_{0})$ is
a \textit{saddle point}. In this case the solution corresponding to the
equilibrium point E is highly unstable and the smallest perturbations will
lead to exponential divergence of the solutions and hence the particles -
from the hypersurface (See Fig.~2). The only (measure zero) exception is
provided by the stable manifold of the point E, represented by the lines AE
and BE, along which particles are attracted towards the brane. An example of
this highly unstable "confinement" is provided by the warping function \cite%
{Gremm}
\begin{equation}
f(l)=-b\ln \cosh (cl),  \label{warpfunction}
\end{equation}%
where $b$ and $c$ are positive constants. Here we have a unique equilibrium
point at $l=0$ and it is easy to verify that $f^{\prime \prime }(0)<0$ in
this case. In this connection we note that for large values of $l$ the
warping function (\ref{warpfunction}) approaches that of the Randall-Sundrum
metric \cite{Randall}) \newline
\[
ds^{2}=e^{-2k\left\vert l\right\vert }\eta _{\alpha \beta }dx^{\alpha
}dx^{\beta }-dl^{2},
\]%
where $k$ is a constant. In this case $f^{\prime }(l)=\mp k$ according to
whether $l$ is positive or negative. Hence for $l\neq 0$ there exist no
equilibrium points, and therefore no confinement of particles purely due to
gravitational effects. However, in this thin wall limit if we take the
extrinsic curvature of the brane $l=0$ as being zero, then (\ref{dynamical1}%
) implies confinement. Nevertheless, as was pointed out in ref. \cite{Muck}
this is a highly unstable confinement in the sense that any small
transversal perturbations in the motion of massive particles along the brane
will cause them to be expelled into the extra dimension. However, this case
lies somewhat outside the scope of the metrics that we are considering here
as in this case the warping function is not smooth (the first derivative of $%
f(l)$\ with respect to the extra coordinate is not continuous at $l=0$). In
the braneworld scenario, as is well known, stable confinement of matter
fields is possible at the quantum level if we take into account interaction
with a scalar field \cite{Rubakov}. In a purely classical picture, however,
one would require mechanisms other than geodesic confinement in order to
constrain massive particles to move on hypersurfaces in a stable way.
\begin{figure}[tbp]
\includegraphics[height=3.0in,width=3.0in]{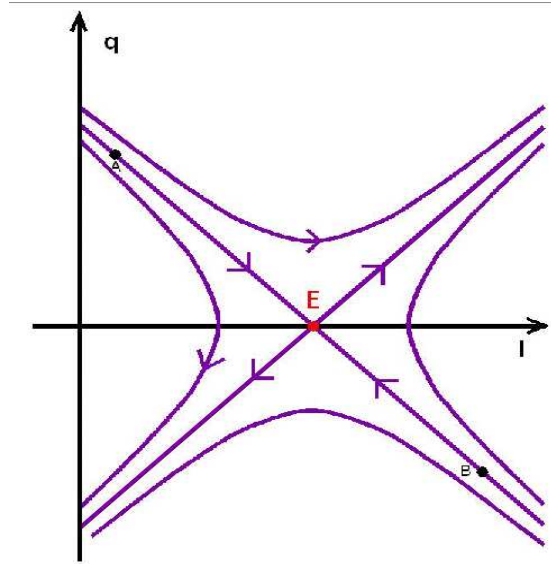}
\caption{When $f^{\prime \prime }(l_{0})<0$ the equilibrium point E is a
saddle. In this case confinement is highly unstable. }
\label{Fig2}
\end{figure}
\newline

Case III. If $f^{\prime \prime }(l_{0})=0$, then both eigenvalues are zero,
which corresponds to a degenerate case. To carry out a qualitative analysis
of the solutions near the equilibrium point, in this case, would require the
knowledge of the third (or higher) derivative of the function $f(l)$. We
shall not consider this case in its generality. However, if $f(l)$ is
constant (case IV), then we can easily draw a global phase portrait of the
system.\emph{\ } \newline

Case IV. If $f(l)$= constant, then $f^{\prime }(l)$ vanishes for all values
of $l$, which implies there is an infinite number of non-isolated
equilibrium points, that is, we have a line of equilibrium points ($q=0$).
Perturbations along this line are neutrally stable, which implies that
particles placed on any of the hypersurfaces of the foliation $l=$ constant,
will remain there so long as they do not receive a transversal velocity to
the hypersurface. \newline

%
%
Case V. If there are no equilibrium points, that is, the warping function $%
f(l)$ does not have any turning points for any value of $l$, then we cannot
have confinement of classical particles to hypersurfaces solely due to
gravitational effects. An example of this situation is illustrated by the
warping function $f(l)=\frac{1}{2}\ln \left( \Lambda l^{2}/3\right) $%
considered in Ref. \cite{Mashhoon}.

We note that even when there are no equilibrium points, the picture of
dynamics can still be obtained if $f(l)$ is known, by obtaining a first
integral of the system (\ref{dynamical}). This can easily be done by writing
this system as the first-order differential equation%
\[
\frac{dl}{dq}=-\frac{q}{f^{\prime }(\epsilon +q^{2})},
\]%
which can be readily integrated to yield%
\begin{equation}
f(l)=-\ln \sqrt{\epsilon +q^{2}}+K,  \label{first-integral}
\end{equation}%
where $K$ is a constant of integration.

\subsection{The 5D motion of photons near the hypersurface\protect\bigskip}

Finally we examine the motion of photons (null geodesics) and ask whether
they can be confined to or a neighbourhood of a hypersurface. In this case
the dynamical system (\ref{dynamical}) becomes
\begin{eqnarray}
\frac{dl}{d\lambda } &=&q  \nonumber  \label{dynamical2} \\
\frac{dq}{d\lambda } &=&-f^{\prime }q^{2}.
\end{eqnarray}%
The equilibrium points now are given by $q=0$, which consists of a line of
equilibrium points along the $l$-axis, with eigenvalues both equal to zero.
As a result there
exist 5D null geodesics in any hypersurface $l=$ constant. This demonstrates
that confinement of photons to hypersurfaces does not depend upon the
warping factor. There are also regions of stability and instability with
respect to small perturbations along the $l$-axis. As was pointed out in the
previous section we can easily obtain a first integral of the system (\ref%
{dynamical2})\ which is given by ( \ref{first-integral}). In the case of the
motion of photons ($\epsilon =0$) this gives
\[
q=Ae^{-f(l)}
\]%
where $A$ is a constant of integration. Therefore we can obtain a global
picture of the solutions (\ref{dynamical2})\ provided that we have some
qualitative knowledge of the function $f(l)$.

\begin{figure}[tbp]
\caption{Depicted in the top panel is the phase portrait representing the
motion of photons in presence of $Z_{2}$ symmetry with an increasing
monotonic function $f(l)$ for $l\geq 0$ and $f(l)\rightarrow \infty $ as $l$
$\rightarrow \infty $. There is confinement of photons at any hypersurface $%
\Sigma $ of the bulk foliation $l=$ constant. The bottom panel shows the
corresponding phase portrait for the cases where $f(l)$ is a decreasing
monotonic function for $l\geq 0$ that approaches the limit $f(l)\rightarrow
-\infty $ as $l\rightarrow \infty $. Photons that are not confined are
repelled by the hypersurface $\Sigma $ $(l=0)$.}
\label{Fig3}\includegraphics[height=3.0in,width=3.0in]{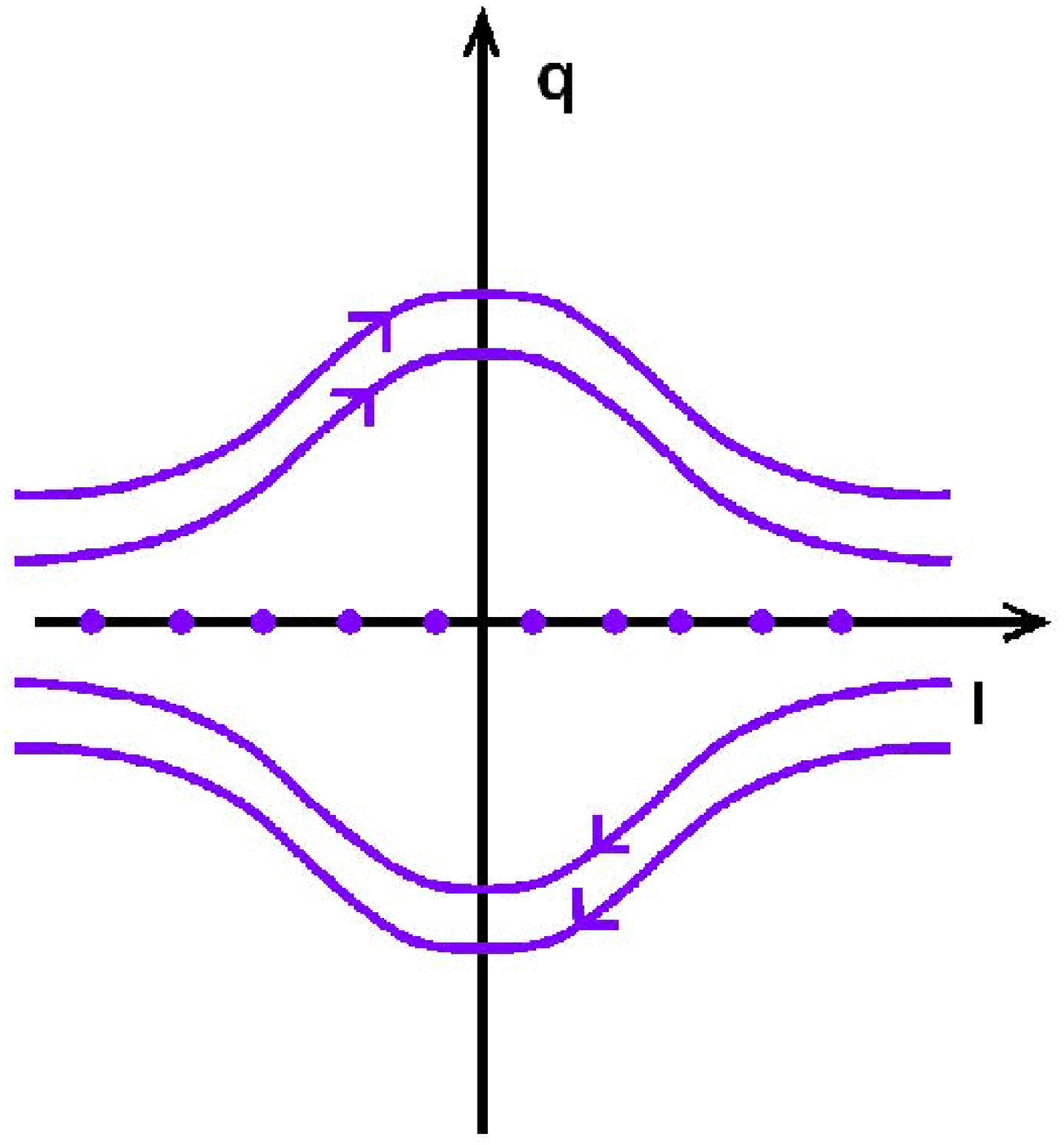} %
\includegraphics[height=3.0in,width=3.0in]{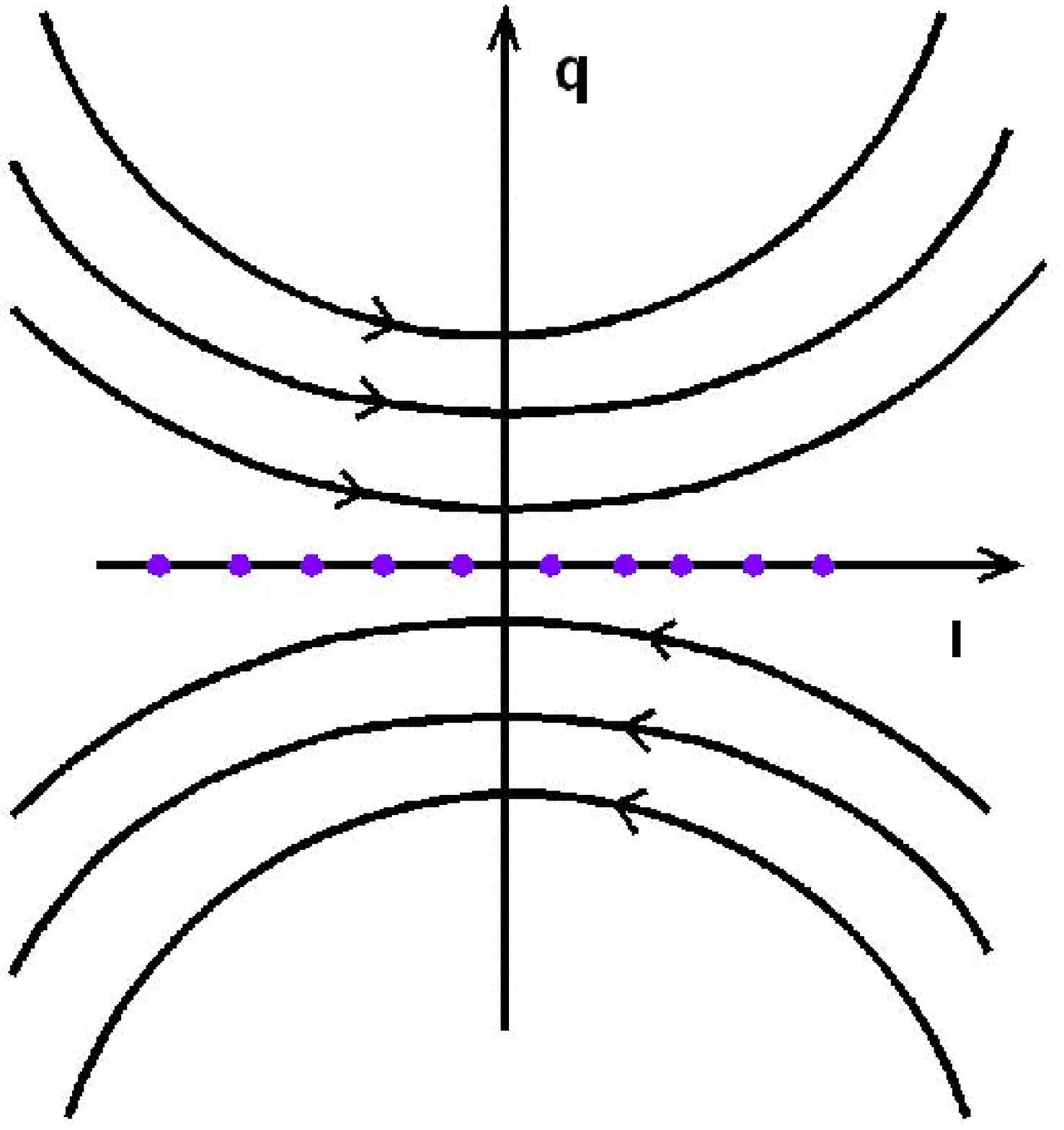}
\end{figure}

Many different cases can arise, depending upon the nature of $f(l)$. Here
for definiteness and simplicity we impose $Z_{2}$ symmetry on the geometry
of the bulk. The following are some possible examples:

\begin{itemize}
\item[(i)] $f(l)$ is an increasing monotonic function for $l\geq 0$ which
approaches the limit $f(l)\rightarrow \infty $ as $l\rightarrow \infty $
(see the top panel of Fig.~\ref{Fig3}). This includes the case of the
warping function considered in (\cite{Mashhoon}).

\item[(ii)] $f(l)$ is a decreasing monotonic function for $l\geq 0$ that
approaches the limit $f(l)\rightarrow -\infty $ as $l\rightarrow \infty $,
which is the case of the warping function given by (\ref{warpfunction}) (see
the lower panel of Fig.~\ref{Fig3}). Note the interesting behaviour of
photons that are not confined: if we set them in motion towards the
hypersurface $\Sigma (l=0)$ then after reaching a minimal distance from $%
\Sigma $ they bounce back to infinity, showing clearly that they are
repelled by $\Sigma $;

\item[(iii)] $f(l)$ is constant, in which case we have the same phase
diagram as in the case of massive particles. 
\end{itemize}

Finally it is worth noting that in the case of null geodesics, the
equilibrium points of the Eqs. (\ref{dynamical2}) do not require $f^{\prime
}=0$, which is a consequence of the fact that the above mentioned theorem
\cite{Eisenhart} does not apply to null geodesics.

\section{Conclusions}

We have studied some aspects of the motion of massive particles and photons
in five-dimensional warped product spaces. Spaces of this type have received
a great deal of attention over the recent years mainly in connection with
the braneworld scenarios. Our treatment has been geometrical and classical
in nature and we have considered confinement to hypersurfaces purely due to
classical gravitational effects.

Employing the splitting that naturally occurs in such spaces between the
motion in the fifth dimension and the remaining dimensions, we have found,
using phase plane analysis, a novel form of \textit{quasi-confinement} which
is neutrally stable. In a sense this amounts to a generalisation of the $%
\delta $ -confinement of particles in thin branes. An important reason to
consider models outside the framework of thin branes is that there is a
characteristic minimum length given by the string scale \cite%
{Seahra,Csaki,Gremm,Wolfe}.

The importance of the type of confinement discussed here is that it is due
purely to the classical gravitational effects, without requiring the
presence of brane type confinement mechanisms.

As an example, we have found that in the case of the Randall-Sundrum metric
gravity alone is not capable of confining massive particles, although
photons may be confined. As is well known, the particle confinement in this
scenario is achieved by means of a scalar field \cite{Randall}.

An assumption usually made in braneworld scenarios, which plays an important
role in connection with confinement in these models, is that of $Z_{2}$
symmetry. It is therefore important to take a closer look at the models we
have found, which are capable of providing successful confinement, from this
perspective.

Recalling that $Z_{2}$ symmetry implies
\begin{equation}
\Omega _{\alpha \beta }^{+}+\Omega _{\alpha \beta }^{-}=0,
\end{equation}%
it is easy to check that this condition is satisfied in cases where the
hypersurfaces are located at the turning points $l=l_{0}$ of the functions $%
f $. This follows from the fact that in such cases $f^{\prime }$, and hence $%
\Omega _{\alpha \beta }=-f^{\prime }e^{2f}h_{\alpha \beta }(x)$, have
different signs on either sides of the hypersurface and thus $Z_{2}$
symmetry is satisfied by default. Importantly this includes all the
equilibrium points we have found in the massive particle case. \newline

Finally it is interesting to compare our results with the classical theorem,
according to which the geodesics of a Riemannian space coincide with the
geodesics of an embedded (totally geodesic) hypersurface if and only if the
extrinsic curvature of that hypersurface vanishes. The analysis involved in
proving this theorem \cite{Eisenhart}, though general, does not address the
case of null geodesics, as it confines itself to positive definite metrics.
Nor does it provide any information regarding the stability of the this
coincidence. Our analysis on the other hand, although restricted to the case
of warped product spaces, provides information regarding the stability of
confinement to totally geodesic hypersurfaces, through the use of dynamical
system analysis.

\section{Acknowledgement}

F. Dahia and C. Romero would like to thank CNPq-FAPESQ (PRONEX) for
financial support. We also thank M. A. S. Cruz for help with the figures.

\end{document}